\documentclass[12pt,a4paper]{article}

\usepackage{latexsym}
\usepackage{wasysym}

\title{Testing a Crucial Assumption}

\author{Ll.\ Bel \footnote{wtpbedil@lg.ehu.es}\\
\emph{Fisika Teorikoa, Euskal Herriko Unibertsitatea}, \\
\emph{P.K. 644, 48080 Bilbo, Spain}
}

\begin{document}
\maketitle

\begin{abstract}

Simplified description of an experiment of the Michelson-Morley type being completed at the University of Western Australia with a discussion of the possible meanings of its outcome. 

\end{abstract}

\section*{Introduction}

Besides the statements of its general principles and main foundational equations every physical theory requires some decreed assumptions to establish a link between physical measurable quantities and its theoretical counterparts. Special and General relativity are no exceptions except for the fact that after so many years and so many accomplishments the physical quantity length and the derived concept of distance has remained elusive to say the least. 

The most explicit of these interpretative assumptions in relativistic theories is that one that declares that the interval of time measured with an atomic clock should be identified with the interval of proper time calculated, for any space-time model, along the world-line describing the motion of the clock.

There is no generally accepted counterpart to this assumption when the concepts of length or distance are involved although the main trend tries to reduce any measure of distance to a measure of time via light signals ignoring thus any reference to a more fitting concept. We show in this paper that this does not need to be so and that other possibilities remain to choose interpretative assumptions for the quantity length that can be falsified by experiments,  and in particular by the Western Australia experiment by Tobar et al., \cite{Tobar}, now in process of being completed\footnote{See also the reference \cite{Stanwix} added in this replacement}.

\section{Frames of reference with constant angular velocity}

Let $ds^2$ be the line element of Minkowski's space-time referred to a system of Cartesian coordinates of a Galilean frame of reference so that:

\begin{equation}
\label {1.1}
ds^2=-c^2dT^2+dX^2+dY^2+dZ^2
\end{equation}
The motion of a frame of reference with constant angular velocity $\Omega$ is a time-like Killing congruence with parametric equations such as:

\begin{eqnarray}
\label {1.2a}
T&=&t \\
\label {1.2b}
X&=&x\cos\,\Omega t-y\sin\,\Omega t \\
\label {1.2c}
Y&=&x\sin\,\Omega t+y\cos\,\Omega t \\
\label {1.2d}
Z&=&z
\end{eqnarray}
where $x,y,z$ are coordinates of the quotient space and the parameter t is a particular time coordinate defining one among all other possible  synchronizations. Using this system of adapted coordinates to the rotating frame of reference the line-element of the space-time becomes:

\begin{equation}
\label {1.3}
ds^2=-\xi^2c^2dt^2+2\omega(xdy-ydx)cdt+dx^2+dy^2+dz^2
\end{equation}
where $\omega=\Omega/c$, and :

\begin{equation}
\label {1.5}
\xi=\sqrt{1-\omega^2\rho^2},\quad  \rho=\sqrt{x^2+y^2}
\end{equation}
that can be split as follows:

\begin{equation}
\label {1.4}
ds^2=-\psi^2+d\hat s^2
\end{equation}
with:

\begin{equation}
\label {1.5a}
\psi=-\xi cdt+\omega\xi^{-1}(xdy-ydx) 
\end{equation}
which is the 1-form whose components are the covariant components of the unit vector field tangent to the killing congruence, and with:

\begin{equation}
\label {1.5b}
d\hat s^2=dx^2+dy^2+dz^2
+\omega^2\xi^{-2}(y^2dx^2+x^2dy^2-2xydxdy)
\end{equation}

Let us consider a domain of the rotating frame of reference where there is at rest a medium with index of refraction $n$ so that the velocity of light in this medium is $v_n=c/n$. According to well established theory, \cite{Synge}, if light is propagating freely it will follow a geodesic of the following space-time metric:

\begin{equation}
\label {1.6}
ds^2_n=ds^2+(1-\frac{1}{n^2})\psi^2
\end{equation}
And similarly if light, because of multiple reflexions, or equivalent constraints, is forced to propagate along any space circuit, still its propagation will have to satisfy the condition:

\begin{equation}
\label {1.7}
ds_n=0.
\end{equation}
Let us consider in particular a closed circuit $C$ with parametric equations:

\begin{equation}
\label {1.8}
x=f_x(\chi), \quad y=f_y(\chi), \quad z=f_z(\chi)
\end{equation}
such that:

\begin{equation}
\label {1.9}
f_x(\chi_0)=f_x(\chi_1)=x_0, \quad f_y(\chi_0)=f_y(\chi_1)=y_0, \quad f_z(\chi_0)=f_z(\chi_1)=z_0 
\end{equation}
From (\ref{1.6}) and (\ref{1.7}) it follows that:

\begin{equation}
\label {1.10}
\frac{dt}{d\chi}=\frac{n}{c}\xi^{-1}\frac{d\hat s}{d\chi}+ \frac{\omega}{c}\xi^{-2}(x\frac{dy}{d\chi}-y\frac{dx}{d\chi}) 
\end{equation} 
The sign of the second term in the r-h-s depends upon the direction in which light is propagating. Therefore if two light rays leave the point $P_0$ at proper time $\tau_0$ and propagate in different directions then they will reach again the point $P_0$ at two different proper times $\tau_+$ and $\tau_-$ such that:

\begin{equation}
\label {1.11}
\frac12(\tau_+-\tau_-)=\frac{\omega}{c}\xi_0\int_{C}\xi^{-2}(xdy-ydx)
\end{equation}
$\xi_0$ being the value of $\xi$ at the point $P_0$.
This result is the celebrated Sagnac effect.

This paper will show the importance of the complementary result:

\begin{equation}
\label {1.12}
\frac12(\tau_++\tau_-)=\frac{n}{c}\xi_0\int_{C}\xi^{-1}d\hat s
\end{equation}  

\section{Free mobility of ideal rigid bodies}

Suppose that we want to give an operational meaning to Eq. (\ref{1.12}).
The l-h-s has a very clear meaning: it is the mean proper period of light propagating in both directions. This is not the case of the r-h-s. To make sense of it we need to be able to identify the measurable properties of the circuit with the geometrical figure that describes it. If in particular we are dealing with a circular circuit $C$ with a measured radius $r$ according to any given protocol we need to know how to use this value using a chosen geometry of space. 
More precisely: we need to know how to derive from this value of $r$ the parametric equations of the circuit $C$, and then to calculate the line-integral of Eq. (\ref{1.12}).  

We know that the result should be intrinsic, i.e. independent of the system of coordinates being used, but it can not be, as we show below independent of a crucial choice of a metric to describe the geometry of space. Is this choice that we claim in Sec.~4 can be subject to an experimental test. Let $dl^2$ be this metric, that at this stage it can be considered as an auxiliary metric, chosen for convenience:

\begin{equation}
\label {2.1.1}
dl^2=a_{ij}dx^idx^j
\end{equation}
where $x^1=x$, $x^2=y$ and $x^3=z$. To give a meaning to the r-h-s of Eq.~ (\ref{1.12}) the strategy that we shall implement in Sec.~4  will be the following: i)~ we shall
obtain and use the geodesic coordinates, \cite{Thomas}, $x^{i^\prime}=x^{i^\prime}(x^j)$ of (\ref{2.1.1}) so that:

\begin{equation}
\label {2.1.2}
a_{{i^\prime}{j^\prime}}=\delta_{{i^\prime}{j^\prime}}
-\frac13 R_{{i^\prime}{k^\prime}{j^\prime}{l^\prime}}x^{k^\prime}x^{l^\prime}
\end{equation}
to the second order of approximation around the point $C_0$, the center of the circle, with coordinates $x^{i^\prime}=0$; 

and ii) we shall use the fact that in this system of coordinates a geodesic disk of radius $r$ centered at $C_0$ is a central section of the sphere:

\begin{equation}
\label {2.1.3}
(x^{1^\prime})^2+(x^{2^\prime})^2+(x^{3^\prime})^2=r^2;
\end{equation}

The most apparently natural choice that we could make to choose a metric of space would be to assume that  $dl^2=d\hat s^2$ because:

\begin{equation}
\label {2.1.4}
\int_L{d\hat s^2}=\int_L{ds^2}
\end{equation}
where $L$ is any path orthogonal to the Killing congruence of the frame of reference. This is the choice that most authors take for granted.

A second choice to describe the geometry of space that is directly suggested by (\ref{1.12}), but has also been proposed following other considerations, \cite{Carter}, is:

\begin{equation}
\label {2.3}
d\check s^2=\xi^{-2}d\hat s^2
\end{equation}

A third choice that we shall consider in Sec.~4 is:

\begin{equation}
\label {2.3.1}
d\tilde s^2=\xi^2d\hat s^2
\end{equation}
This metric has been considered in \cite{Fock}, \cite{Bel69} and  \cite{Bel71} because it cures some odd anomalies of the Schwarzschild and Curzon space-times, and for technical convenience in \cite{Geroch} and \cite{Martin}.

But none of these metrics is flat and this poses a serious problem of principle.  In fact it is elementary knowledge of Riemannian geometry that whatever is this geometry, as long as it does not has  constant curvature, the ratio $L/r$, L being the perimeter of a geodesic circle, depends on the location and the orientation of this circle. That is alright to model a particular circular circuit with a measured physical radius $r$ and perimeter $L$ as long as we restrain from moving it from point to point or from changing its orientation. But, in principle, no metrologist should accept these restrictions. 

In Sect.~4 we shall see that under the conditions of the Western Australia experiment we may be tempted to dismiss this free mobility problem of ideal rigid bodies as irrelevant, and this would be indeed the case if we thought only in terms of manufacturing specifications of real rigid bodies. But as we shall see the prediction of the outcome of the experiment is very sensitive to whatever choice is made: either to dismiss it or to cure the problem. 

A fourth choice that we shall consider is to implement the Principle of free mobility introducing a flat metric\footnote{A constant non zero curvature does not seem appropriate here.} intrinsically associated to the Killing congruence. We developed elsewhere, \cite{Bel96}, a general theory to do that but the full theory is not needed here because the metric (\ref{1.5b}) is particularly simple and can be written as:

\begin{equation}
\label {2.4}
d\hat s^2=d\kappa^2+dz^2
\end{equation}  
where $d\kappa^2$ is a 2-dimensional Riemannian metric of a 2-dimensional manifold with coordinates $x,y$. Therefore we know from one of Gauss's theorems that it is 
possible to find a function $\mu$ such that the metric:

\begin{equation}
\label {2.5}
d\bar s^2=e^{2\mu}d\kappa^2+dz^2
\end{equation} 
be flat. Actually the function $\mu$ that satisfies the required regularity conditions at the axis of symmetry is:

\begin{equation}
\label {2.6}
\mu=\xi-\ln(1+\xi^{-1})-1+\ln\,2
\end{equation} 
and it is unique.

\section{Whispering gallery modes}

Whispering gallery modes are being used in the Michelson-Morley experiment that is in the process of completion  at the University of Western Australia and that we shall analyze in the next section from the theoretical point of view. Below we shall give a very simplified description of what these modes are but we refer interested readers to references \cite{Tobar}, \cite{Wolf}.

Let us consider a circular ring, made of a dielectric with magnetic permeability $\mu=1$ and electric permittivity $\epsilon$ so that the index of refraction is $n=c\sqrt{\epsilon}$, where light circulates around in both directions. We assume that the amplitude $A$ of either wave is the same as well as the frequency $\nu$ but, although as we shall see in the next section this will be irrelevant, for completeness we do not assume that the periods of rotation $\tau_+$ and $\tau_-$ are equal. Consequently we define two wave-lengths $\lambda_+$ and $\lambda_-$ such that:

\begin{equation}
\label {3.1}
\nu\lambda_+=\frac{L}{n\tau_+}, \quad \nu\lambda_-=\frac{L}{n\tau_-} 
\end{equation}   
where $L$ is the length of the ring. The amplitude $\Phi$ of the superposition of the two waves is:

\begin{equation}
\label {3.2}
\Phi=A\cos(2\pi\nu t-2\pi\frac{l}{\lambda_+})+
A\cos(2\pi\nu t+2\pi\frac{l}{\lambda_-})
\end{equation}
where $l$ is the winding length around the circle measured from an arbitrary point on it. Equivalently we can write:

\begin{equation}
\label {3.3}
\Phi=A\cos(2\pi\nu t)
\left(\cos\,2\pi\frac{l}{\lambda_+}+ \cos\,2\pi\frac{l}{\lambda_-}\right)+
A\sin(2\pi\nu t)
\left(\sin\,2\pi\frac{l}{\lambda_+}- \sin\,2\pi\frac{l}{\lambda_-}\right)
\end{equation}

Zero vales of $\Phi$ at fixed angular positions independent of time, i.e. nodes, will set up whenever $l$ is a solution of the system of equations:

\begin{eqnarray}
\label {3.4}
\cos\,2\pi\frac{l}{\lambda_+}+ \cos\,2\pi\frac{l}{\lambda_-}&=&0 \\
\sin\,2\pi\frac{l}{\lambda_+}- \sin\,2\pi\frac{l}{\lambda_-}&=&0
\end{eqnarray}
This will happen when:
 
\begin{equation}
\label {3.5}
l=\left(k+\frac12\right)\frac{\lambda}{2} 
\end{equation}
where : 

\begin{equation}
\label {3.5.1}
\lambda=2\frac{\lambda_+ \lambda_-}{\lambda_++\lambda_-}
\end{equation}
is a reduced wave-length, and $k$ is any integer.

The condition to have a resonant state, i.e. standing waves with a finite number of nodes is that there exist an integer $N$, the number of nodes such that :

\begin{equation}
\label {3.6}
N\frac{\lambda}{2}=L
\end{equation}  

From (\ref{3.1}) we have:

\begin{equation}
\label {3.7}
\lambda_++\lambda_-=
\frac{L}{n\nu}\,\frac{\tau_++\tau_-}{\tau_+\tau_-} 
\quad \hbox{and}
\quad \lambda_+\lambda_-=
\left(\frac{L}{n\nu}\right)^2\,\frac{1}{\tau_+\tau_-}
\end{equation} 
and therefore from (\ref{3.5}) we get:

\begin{equation}
\label {3.8}
\lambda=\frac{2L}{n\nu(\tau_++\tau_-)}
\end{equation}
and, finally, from (\ref{3.6}) we obtain the useful formula:

\begin{equation}
\label {3.9}
\nu=\frac{N}{n(\tau_++\tau_-)}
\end{equation}

\section{The Western Australia experiment}

In the Western Australia experiment two Whispering gallery modes are excited around two great circles of the surface of two sapphire spheres with a diameter of 5\,cm. One of them, say {\bf A}, has one of its diameters in the direction of the vertical of the location and it is slowly rotating around this direction. The second one, say {\bf B}, is horizontal and although it rotates with {\bf A} we may as well consider it as fixed for the purpose of analyzing the experiment. The frequencies $\nu_A$ and $\nu_B$ are continuously monitored and compared. We give below our analysis of this experiment.

The experiment is being made at the University of Western Australia that is at a co-latitude $\theta=58^\circ$ South. Therefore we have:

\begin{equation}
\label {4.2}
\delta\equiv \left(\frac{\Omega\rho_0}{c}\right)^2\approx 1.7\times 10^{-12}, \quad \rho_0=R\sin\,\theta
\end{equation}
where $R$ is the radius and $\Omega$ is the angular velocity of the Earth, which is the relevant frame of reference to consider here. From now on we shall neglect the powers of $\delta$. 
     
At this approximation we have the following expressions for the four metrics (\ref{2.3})-(\ref{2.5}):

\begin{eqnarray}
\label {4.5}
d\hat s^2=(1+\omega^2y^2)dx^2+(1+\omega^2x^2)dy^2-2\omega^2xydxdy+dz^2\\
d\check{s}^2=(1+\omega^2(2y^2+x^2))dx^2+(1+\omega^2(2x^2+y^2))dx^2 \nonumber \\
-2\omega^2xydxdy+(1+\omega^2\rho^2)dz^2 \\
d\tilde{s}^2=(1-\omega^2x^2)dx^2+(1-\omega^2y^2)dx^2-2\omega^2xydxdy+
(1-\omega^2\rho^2)dz^2\\
d\bar s^2=(1-\frac12\omega^2(3x^2+y^2))dx^2
+(1-\frac12\omega^2(3y^2+x^2))dy^2 \nonumber \\-2\omega^2xydxdy+dz^2
\end{eqnarray}  

Since by construction the fourth metric $d\bar s^2$ is Euclidean there exists a system of coordinates that will bring it to the Cartesian form:   

\begin{equation}
\label {4.6}
d\bar s^2=d\bar x^2+d\bar y^2+d\bar z^2
\end{equation}
To do that the appropriate change of coordinates is:

\begin{equation}
\label {4.7}
x=\bar x+\frac14\omega^2\bar x\bar\rho^2,\quad 
y=\bar y+\frac14\omega^2\bar y\bar\rho^2, \quad z=\bar z
\end{equation}
Using this convenient system of coordinates the four metrics above can be written collectively as: 

\begin{equation}
\label {4.8}
dl^2=(1+\alpha\omega^2\bar\rho^2)(d\bar x^2+d\bar y^2)+
(1+\beta\omega^2\bar\rho^2)d\bar z^2
\end{equation}
where $\alpha$ and $\beta$ are two parameters whose values are:

\begin{eqnarray}
\label {4.9}
\hbox{ if }dl^2&=&d\bar s^2 \hbox{ then }\alpha=\ \ \ 0, \quad \beta=\ 0 \\
\hbox{ if }dl^2&=&d\hat s^2 \hbox{ then }\alpha=3/2, \quad \beta\ =0 \\
\hbox{ if }dl^2&=&d\check s^2 \hbox{ then }\alpha=5/2, \quad \beta=+1 \\
\hbox{ if }dl^2&=&d\tilde s^2 \hbox{ then }\alpha=1/2, \quad \beta=-1 
\end{eqnarray} 

The non zero strict components of the Riemann tensor of $dl^2$ are:

\begin{equation}
\label {4.10}
R_{1212}=-2\alpha\omega^2, \quad R_{3131}=R_{2323}=-\beta\omega^2 
\end{equation}

Let us now name $\bar x_0$, $\bar y_0$ and $\bar z_0$  the coordinates of the location of the experiment with: 

\begin{equation}
\label {4.18}
\bar x_0=R\sin\,\theta\cos\,\phi, \quad \bar y_0=R\sin\,\theta\sin\,\phi \quad 
\bar z_0=R\cos\,\theta
\end{equation}

The metric $dl^2$ in a geodesic system of coordinates anchored at the point $ C_0$ with coordinates $x^\prime=y^\prime= z^\prime=0$ can be derived using (\ref{2.1.2}) and (\ref{4.10}). We get thus:   

\begin{eqnarray}
\label {4.19}
dl^2=(1+\frac13(2\alpha\omega^2 y^{\prime 2}+\beta\omega^2 z^{\prime 2}))dx^{\prime 2}+ 
(1+\frac13(2\alpha\omega^2 x^{\prime 2}+\beta\omega^2 z^{\prime 2}))dy^{\prime 2}\nonumber\\
+(1+\frac13\beta\omega^2\rho^{\prime 2})dz^{\prime 2}
-\frac43\alpha\omega^2 x^\prime y^\prime dx^\prime dy^\prime -\frac23\beta\omega^2(x^\prime dx^\prime +y^\prime dy^\prime )z^\prime dz^\prime 
\end{eqnarray}
The coordinate transformation that brings $dl^2$ to this form is:

\begin{eqnarray}
\label {4.20}
\bar x=\bar x_0+ x^\prime-\frac12\alpha\omega^2(\bar\rho_0^2x^\prime+\bar x_0(x^{\prime 2}-y^{\prime 2})+2\bar y_0x^\prime y^\prime +\frac13 x^\prime\rho^{\prime 2})\nonumber\\
+\frac12\beta\omega^2z^{\prime 2}(\bar x_0+\frac13 x^\prime) \\
\bar y=\bar y_0+ y^\prime-\frac12\alpha\omega^2(\bar\rho_0^2y^\prime+\bar y_0(y^{\prime 2}-x^{\prime 2})+2\bar x_0y^\prime x^\prime +\frac13 y^\prime\rho^{\prime 2})\nonumber\\
+\frac12\beta\omega^2z^{\prime 2}(\bar y_0+\frac13 y^\prime)
\\ 
\bar z=\bar z_0+ z^\prime-\beta\omega^2z^\prime(\frac12\bar\rho_0^2+\bar x_0x^\prime+\bar y_0y^\prime+\frac13\rho^{\prime 2})
\end{eqnarray}

Let us consider a geodesic circle with center at $C_0$ lying in a plane orthogonal to the axis of rotation. The parametric equations will be then:  

\begin{equation}
\label {4.14}
x^\prime=r\cos\chi, \quad  y^\prime=r\sin\chi, \quad z^\prime=0;
\end{equation} 
where $r$ is its geodesic radius, i.e. its proper length in the sense of the chosen geometry of space. Calculating the perimeter of the circle we get:

\begin{equation}
\label {4.15}
L_h=\int_C{dl}=2\pi r(1+\frac13\alpha\omega^2 r^2)
\end{equation} 
While if the plane of the circle contains the axis of rotation  then its parametric equations are for example:

\begin{equation}
\label {4.16}
x^\prime=r\cos\chi,  \quad y^\prime=0, \quad  z^\prime=r\sin\chi
\end{equation}
and we get:

\begin{equation}
\label {4.17}
L_v=\int_C{dl}=2\pi r(1+\frac16\beta\omega^2 r^2)
\end{equation}  
This shows that the ratio $L/r$ depends, strictly speaking, upon the orientation of the circle, except in the case when $dl^2=d\bar s^2$, i.e. $\alpha=\beta=0$. Therefore none of the remaining metrics should be used in principle to describe the geometrical properties of ideal rigid bodies. On the other hand since:

\begin{equation}
\label {4.17.1}
\eta\equiv\frac{\Omega r}{c}\approx 10^{-15}
\end{equation}
it can be argued that this change of shape is well beyond the specifications under which the sapphire spheres are manufactured and therefore are irrelevant to the experiment that we are describing. This would be to forget that the modifications shown by Eqs. (\ref{4.15}) and (\ref{4.17}) are not elastic deformations i.e.: they do not depend on any physical property. In fact in our opinion they may just reveal an inadequacy in the choice of metric. The purpose of this paper is to claim that these two points of view can be tested by the Western Australia experiment.

From (\ref{1.12}) and (\ref{3.9}) it follows that:

\begin{equation}
\label {4.1}
\nu=\frac{Nc}{2n^2}\left(\xi_0\int_{C}\xi^{-1}d\hat s\right)^{-1}
\end{equation}
a formula where we can now calculate the r-h-s making different choices of geometries and choosing $C$ to be in succession the circuit of the whispering gallery mode $\bf A$ and $\bf B$. Using a system of geodesic coordinates of $dl^2$, it follows from (\ref{4.20}) that the line-element $d\check s^2$, i.e. the integrand in (\ref{4.1}), becomes, neglecting any terms of order $\delta\eta$ or smaller and identifying $\bar\rho_0$ with $rho_0$ :  

\begin{equation}
\label {4.21}
d\check s^2=(1+(5/2-\alpha)\omega^2\rho_0^2)(dx^{\prime 2}+dy^{\prime 2})
+(1+(1-\beta)\omega^2\rho_0^2)dz^{\prime 2}
\end{equation} 
therefrom, choosing appropriately the parametric equations of the circuit $C$ the r-h-s of (\ref{4.1}) can be calculated.

Let us choose as origin of the angle $\varphi$ describing the rotation of {\bf A} 
the position of the plane of this circle when it lyes along the meridian of {$C_0$}. The parametric equations of the circle {\bf A} are then:

\begin{eqnarray} 
\label {4.22}
x^\prime &=&r(\cos\,\varphi\cos\,\chi\cos\,\theta-\sin\,\chi\sin\,\theta) \\
y^\prime &=&r\sin\,\varphi\cos\,\chi \\
z^\prime &=&r(\sin\,\chi\cos\,\theta+\cos\,\varphi\cos\,\chi\sin\,\theta) 
\end{eqnarray}  
and those of {\bf B} are:

\begin{eqnarray}
\label {4.23}
x^\prime &=&r(\cos\,\varphi\cos\,\chi-\sin\,\chi\sin\,\varphi)\cos\,\theta \\
y^\prime &=&r(\sin\,\varphi\cos\,\chi+\cos\,\varphi\sin\,\chi) \\
z^\prime &=&r(\cos\,\varphi\cos\,\chi-\sin\,\chi\sin\,\varphi)\sin\,\theta
\end{eqnarray}
with $\chi \in [0,2\pi]$.

We obtain thus for the whispering gallery mode {\bf B} of reference:

\begin{eqnarray}
\label {4.27}
\nu_B=\frac{Nc}{2n^2L}\left(1-\frac14(\frac32-\alpha)\omega^2\rho_0^2(1+\cos^2\,\theta)\right)\nonumber\\
+\frac{Nc}{8n^2L}\beta\omega^2\rho_0^2\sin^2\,\theta
\end{eqnarray}
and for {\bf A} is:

\begin{eqnarray}
\label {4.26}
\nu_A=\frac{Nc}{2n^2L}\left(1-\frac14(\frac32-\alpha)\omega^2\rho_0^2(1+\frac12\sin^2\,\theta(1-\cos\,2\varphi)\right) \nonumber\\
+\frac{Nc}{8n^2L}\beta\omega^2\rho_0^2\left(1-\frac12\sin^2\,\theta(1-\cos\,2\varphi)\right)
\end{eqnarray}
The full signature of the experiment depends on the colatitude $\theta$ and the two parameters $\alpha$ and $\beta$, but special interest deserves the first harmonic of the $\varphi$ dependence of $\nu_A$ at twice the rotation frequency of the platform:

\begin{equation}
\label {4.28}
a_2=\frac18(\frac32-\alpha+\beta)\omega^2 R^2\sin^4\,\theta
\end{equation}

More explicitly when $d\bar s^2$ is chosen as geometry of space then the prediction is:

\begin{equation}
\label {4.29}
a_2\approx 2.3\times 10^{-13} 
\end{equation} 
while it is:

\begin{equation}
\label {4.29.1}
a_2=0 
\end{equation}
for the other three line-elements $dl^2$ that we have considered in (\ref{4.9}). This common zero value comes from the fact that these three metrics are conformal.

Although irrelevant in the analysis of the Western Australia experiment let us mention the order of magnitude of the Sagnac effect in this experiment. Defining:

\begin{equation}
\label {4.30}
S\equiv\frac{\tau_+-\tau_-}{\tau_++\tau_-}
\end{equation}
using (\ref{1.11}) and (\ref{1.12}) at the lowest order of approximation, a calculation similar to the preceding one yields:

\begin{equation}
\label {4.31}
S_B=\frac{\eta}{n}\cos\,\theta, \quad 
S_A=\frac{\eta}{n}\sin\,\phi\sin\,\theta
\end{equation}
with $\eta$ defined in (\ref{4.17.1})

\section*{Conclusion}

The Western Australia experiment has been conceived as a test of Special relativity, but as we understand it, it is mainly a test of a crucial collateral assumption having to do with the geometry of space in a frame of reference co-rotating with the Earth. This is not a trivial matter that can be dismissed invoking a Principle of Lorentz Local Invariance saying that only the instantaneous velocity with respect to a local non-rotating frame of reference is what matters in describing the experiment, because this principle is also a collateral assumption that needs to be substantiated. 

Keeping in mind that the reliability of our analysis strongly depends upon the fidelity of the over-simplified model of Whispering Gallery Modes that we have used to real ones, we give below a short description of possible outcomes with an indication of what they would mean, assuming ideally that the values of the two parameters $\alpha$ and $\beta$ could be tested.

Each couple of values corresponds to a geometry of space, or in other words, to a choice to identify an operational measure of length with a calculated geometrical quantity. We have mentioned four such choices, with different justifications, but we could as well leave the choice open and let experiments decide which is the best choice to make.

The first choice, which in our opinion deserves a particular attention is ($\alpha=0,\ \beta=0$), i.e. $dl^2=d\bar s^2$. If this is correct this will mean that the Principle of Free Mobility of ideal rigid bodies holds and this would open the door to a better understanding of what length and distance mean in Special relativity beyond the Galilean frames of reference, as well as in General relativity. On the other hand if this principle does not hold then we are afraid that it will be harder than we thought to develop a satisfactory general theory about frames of reference.

The three other metrics that we have considered will look reassuring to those who believe in the indiscriminate validity of the Principle of Local Lorentz Invariance because if anyone of these metrics is the correct choice the experiment will not give any meaningful signal: both frequencies $\nu_A$ and $\nu_B$ will remain constant up to the precision that it is achieved. A value of $a_2$ significantly below $10^{-13}$ would undoubtedly favor this interpretation, and could be considered an important conclusion because it would be the first test of the crucial assumption that it implies.

An outcome of the same order of magnitude as in the preceding case with different values of $\alpha$ and $\beta$ would be an encouraging surprise strongly suggesting the dynamical importance of the rotation of the Earth to understand the Western Australia experiment.

Finally, let us say that any other outcome would be a puzzling and exciting surprise, and it is only in this case that we could allow any credibility to the fact that we have peeped beyond the domain of validity of Special relativity.

\section{Acknowledgments} 

I gratefully acknowledge the hospitality of the Theoretical Physics Department of the UPV/EHU, and a careful reading of the manuscript by A. Molina.


\begin{thebibliography}{99}

\bibitem{Tobar} M.~E.~Tobar, J.~G.~Hartnett, J.~D.~Anstie,{\em Phys. Lett. A.}, \textbf{300}, 33 (2002)

\bibitem{Stanwix} P.L. Stanwix, M.E. Tobar, J. Winterflood, E.N. Ivanov, M. Susli, J.G. Hartnett, F. van Kann, P. Wolf, to be published in Proc 2004 IEEE International Ultrasonics, Ferroelectrics, and Frequency Control 50th Anniversary Joint Conference.

\bibitem{Wolf} P.~Wolf, M.~E.~Tobar, S.~Bize, A.~Clairon, A.~N.~Luiten, and G.~Santarelli, {\em arXiv:gr-qc/0401017 }

\bibitem{Synge} J. L. Synge, {\em Relativity: The General Theory}, Ch. XI, {\S}
3, North Holland (1960)

\bibitem{Thomas} T. Y. Thomas, {\em Concepts from Tensor Analysis and Differential Geometry }, Academic Press (1961)

\bibitem{Carter} B.~Carter, M.~A.~Abramowicz and J.~P.~Lasota, {\em Gen. Rel. and Grav.}, \textbf{20}, 11, 1173 (1988)

\bibitem{Fock} V.~Fock, {\em The Theory of Space, Time and Gravitation}, Chap. V, {\S} 55, Pergamon Press (1964)

\bibitem{Bel69} L.\ Bel, {\em Journ.\ Math.\ Phys.}, \textbf{10}, 1601 (1969)

\bibitem{Bel71} L.\ Bel, {\em Gen.\ Rel.\ and Grav.}, \textbf{1}, 337 (1971)

\bibitem{Geroch} R.\ Geroch, {\em Journ.\ Math.\ Phys.}, \textbf{28}, 9, p. 918 (1971)

\bibitem{Martin} J.\ Mart\'{\i}n in {\it Relativity and Gravitation in General}, J.\ Mart\'{\i}n, E.\ Ruiz, F.\ Atrio and M.\ Molina Eds. World Scientific (1999)

\bibitem{Bel96} Ll.\ Bel, {\em Gen.\ Rel.\ and Grav.}, \textbf{28}, 1139 (1996)

\bibitem{Helmholtz} Quoted in D. Laugwitz, {\em Differential and Riemannian
Geometry,} Chap. IV, Academic Press  (1965)

\bibitem{Poincare} H. Poincar\'e {\em La science et l'hypoth\`ese,}
Chap. IV. Any edition

\bibitem{Cartan} E. Cartan {\em G\'eometrie des Spaces de Riemann,}
Chap. V, {\S} III and Chap. VI, {\S} IX. Gauthiers-Villars (1951)



\end{thebibliography}
\end{document}